\documentclass[epj]{svjour}
\renewcommand{\arraystretch}{2}
\usepackage{graphics}
\usepackage{amsmath}
\usepackage{color}

\begin{document}
\title{Atom chips in the real world: the effects of wire corrugation}
\author{T. Schumm\inst{1}, J. Est\`{e}ve\inst{1},  C. Figl\inst{1}\thanks{\emph{Present address:
Universit\"at Hannover, D 30167 Hannover, Germany}},
J.-B.~Trebbia\inst{1}, C. Aussibal\inst{1}, H. Nguyen\inst{1}, D.
Mailly\inst{2}, I. Bouchoule\inst{1}, C.~I.~Westbrook\inst{1} and
A.~Aspect\inst{1}} \institute{Laboratoire Charles Fabry de
l'Institut d'Optique, UMR 8501 du CNRS, 91403 Orsay Cedex, France
\and Laboratoire de Photonique et de Nanostructures, UPR 20 du
CNRS, 91460 Marcoussis, France }

\abstract{We present a detailed model describing the effects of
wire corrugation on the trapping potential experienced by a cloud
of atoms above a current carrying micro wire. We calculate the
distortion of the current distribution due to corrugation and then
derive the corresponding roughness in the magnetic field above the
wire. Scaling laws are derived for the roughness as a function of
height above a ribbon shaped wire. We also present experimental
data on micro wire traps using cold atoms which complement some
previously published measurements~\cite{Nous-fragm} and which
demonstrate that wire corrugation can satisfactorily explain our
observations of atom cloud fragmentation above electroplated gold
wires. Finally, we present measurements of the corrugation of new
wires fabricated by electron beam lithography and evaporation of
gold. These wires appear to be substantially smoother than
electroplated wires. 
    }

\authorrunning{T.~Schumm et al.}
\titlerunning{Atom chips in the real world: the effects of wire corrugation}

\maketitle

\section{Introduction}
\label{intro}

Magnetic traps created by current carrying micro wires have proven
to be a powerful alternative to standard trapping schemes in
experiments with cold atoms and Bose-Einstein
condensates~\cite{Folmanrevue}.
%~\cite{JacobBEC,SchmiedmayerY,BecZimmermann,Hinds2wire,BECmicroKetterle,Mull99,Vuletic-VdW}.
These so-called "atom chips" combine robustness, simplicity and
low power consumption with strong confinement and high flexibility
in the design of the trapping geometry. Integrated atom optics
elements such as waveguides and atom interferometers have been
proposed and could possibly be integrated on a single chip using
fabrication techniques known from microelectronics. Quantum
information processing with a single atom in a micro trap has also
been proposed~\cite{calarco:2000}.

Real world limitations of atom chip performance are thus of great
interest. Losses and heating of atoms due to thermally exited
currents inside conducting materials composing the chip were
predicted theoretically~\cite{Henkel-applphys99,Henk99} and
observed experimentally soon after the first experimental
realizations of atomic micro
traps~\cite{Hinds-heating2003,Cornell-spinfliplosses}.

An unexpected problem in the use of atom chips was the observation
of a fragmentation of cold atomic clouds in magnetic micro
traps~\cite{Zimmermann-fragPRA2002,BECmicroKetterle}. Experiments
have shown that this fragmentation is due to a time independent
roughness in the magnetic trapping potential created by a
distortion of the current flow inside the micro
wire~\cite{Zimmermann-frag2002}. It has also been demonstrated
that the amplitude of this roughness increases as the trap center
is moved closer to the micro wire~\cite{Hinds-frag2003}.
Fragmentation has been observed on atom chips built by different
micro fabrication processes using gold~\cite{Nous-fragm} and
copper wires~\cite{Zimmermann-fragPRA2002,BECmicroKetterle}, and
on more macroscopic systems based on cylindrical copper wires
covered with aluminum~\cite{Hinds-frag2003} and micro machined
silver foil~\cite{vale:2004}. The origin of the current distortion
inside the wires causing the potential roughness is still not
known for every system.

In a recent letter~\cite{Nous-fragm}, we experimentally
demonstrated that wire edge corrugation explains the observed
potential roughness (as theoretically proposed
in~\cite{Lukin-frag2003}) in at least one particular realization
of a micro trap. In this paper, we will expand on our previous
work giving a more detailed description of the necessary
calculations as well as presenting a more complete set of
experimental observations. We emphasize that extreme care has to
be taken when fabricating atom chips, and that high quality
measurements are necessary to evaluate their flatness in the
frequency range of interest. We will discuss the influence of
corrugations both on the edges as well as on the surface of the
wire and give scaling laws for the important geometrical
quantities like atom wire separation and wire dimensions. We will
also present preliminary measurements on wires using improved
fabrication techniques.

The paper is organized as follows. In section~\ref{sec:magtrap},
we give a brief introduction to magnetic wire traps and emphasize
that the potential roughness is created by a spatially fluctuating
magnetic field component parallel to the wire. In
section~\ref{sec:magfield}, we give a general framework to
calculate the rough potential created by any current distortion in
the wire. A detailed calculation of the current flow distortion
due to edge and surface corrugations on a rectangular wire is
presented in section~\ref{sec:j_distortion}. In
section~\ref{sec:flatwire}, we apply these calculations to the
geometry of a flat wire, widely used in experiments. Edge and
surface effects are compared for different heights above the wire
and we present important scaling laws that determine the optimal
wire size for a given fabrication quality. In
section~\ref{sec:experiment} and~\ref{sec:evapwire}, we show
measurements of the spectra of edge and surface fluctuations for
two types of wires produced by different micro fabrication
methods: optical lithography followed by gold electroplating and
direct electron beam lithography followed by gold evaporation. We
also present measurements of the rough potential created by a wire
of the first type using cold trapped Rubidium atoms.

\section{Magnetic micro traps}
\label{sec:magtrap}The building block of atom chip setups is the
so-called side wire guide~\cite{Folmanrevue}. The magnetic field
created by a straight current carrying conductor along the $z$
axis combined with a homogeneous bias field $B_0$ perpendicular to
the wire creates a two-dimensional trapping potential along the
wire (see figure~(\ref{fig.wireandaxes})). The total magnetic
field cancels on a line located at a distance $x$ from the wire
and atoms in a low field seeking state are trapped around this
minimum. For an infinitely long and thin wire, the trap is located
at a distance $x=\mu_0  \, I / (2\, \pi \, B_0)$. To first order,
the magnetic field is a linear quadrupole around its minimum. If
the atomic spin follows adiabatically the direction of the
magnetic field, the magnetic potential seen by the atoms is
proportional to the magnitude of the magnetic field. Consequently,
the potential of the side wire guide grows linearly from zero with
a gradient $B_0/x$ as the distance from the position of the
minimum increases.
%\begin{equation}
%\frac{\partial |B(0,h,0)|}{\partial x} = \frac{\partial
%|B(0,h,0)|}{\partial y} = \frac{\mu_0 I}{2 \pi h^2} =
%\frac{B_{Bias}}{h} \, .
% \label{gradient_straight_wire}
%\end{equation}

For a straight wire along $z$, all magnetic field vectors are in
the $(x,y)$ plane. Three dimensional trapping can be obtained by
adding a spatially varying magnetic field component $B_z$ along
the wire. This can be done by bending the wire, so that a magnetic
field component along the central part of the wire is created
using the same current. Alternatively, separate chip wires or even
macroscopic coils can be used to provide trapping in the third
dimension.

%(Care has to be taken to avoid a vanishing magnetic field at the
%trap center, which would create Spin flip losses)

For a realistic description of the potential created by a micro
wire, its finite size has to be taken into account. Because of
finite size effects, the magnetic field does not diverge but
reaches a finite value at the wire surface. For a square shaped
wire of height and width $a$ carrying a current $I$, the magnetic
field saturates at a value proportional to $I/a$, the gradient
reaches a value proportional to $I/a^2$. Assuming a simple model
of heat dissipation, where one of the wire surfaces is in contact
with a heat reservoir at constant temperature, one finds the
maximal applicable current to be proportional to $a^{3/2}$
\cite{Schmiedmayerheating}. Therefore, the maximal gradient that
can be achieved is proportional to $1/\sqrt{a}$. This shows that
bringing atoms closer to smaller wires carrying smaller currents
still increases the magnetic confinement, which is the main
motivation for miniaturizing the trapping structures. However the
magnetic field roughness arising from inhomogeneities in the
current density inside the wire also increases as atoms get closer
to the wire. This increase of potential roughness may prevent the
achievement of high confinement since the trap may become too
corrugated.

We emphasize that only the $z$ component of the magnetic field is
relevant to the potential roughness. A variation of the magnetic
field in the $(x,y)$ plane will cause a negligible displacement of
the trap center, whereas a varying magnetic field component $B_z$
modifies the longitudinal trapping potential, creating local
minima in the overall potential \cite{Nous-fragm}.

\section{Calculation of the rough magnetic field created by a
distorted current flow in a wire}\label{sec:magfield} In this
section, we present a general calculation of the extra magnetic
field due to distortions in the current flow creating the trapping
potential. By $\vec{j}$ we denote the current density that
characterizes the distortion in the current flow. The total
current density $\vec{J}$ is equal to the sum of $\vec{j}$ and the
undisturbed flow $j_0 \, \vec{e}_z$. As the longitudinal potential
seen by the atoms is proportional to the $z$ component of the
magnetic field, we restrict our calculation to this component. We
thus have to determine the $x$ and $y$ components of the vector
potential $\mathbf{A}$ from which the magnetic field derives. In
the following, we consider the Fourier transform of all the
quantities of interest along the $z$ axis which we define by
\begin{equation}
A_{l,k}(x,y)  =  \frac{1}{\sqrt{2 \, \pi \, L}} \int A_l(x,y,z) \,
e^{-i \, k \, z} \ dz \ ,
\end{equation}
where we have used the vector potential as an example and $l$
stands for $x$ or $y$, $L$ being the length of the wire. We choose
this definition so that the power spectral density of a quantity
coincides with the mean square of its Fourier transform :
\begin{equation}
    \frac{1}{2 \pi} \int e^{i \, k
    \, z} \ \langle A_{l}(z) \, A_{l} (0) \rangle \ dz = \langle |A_{l,k}|^2
    \rangle \ .
\end{equation}

The vector potential satisfies a Poisson equation with a source
term proportional to the current density in the wire. Thus the
Fourier component $A_{l,k}$ satisfies the following time
independent heat equation
\begin{equation}
 \left( \frac{\partial}{\partial x^2} + \frac{\partial}{\partial y^2} \right) A_{l,k}
 - k^2 \, A_{l,k} = -\mu_0 \, j_{l,k} \ .
 \label{eq.A2D}
\end{equation}
where $j_l$ is one component of the current density $\vec{j}$. In
the following, we use cylindrical coordinates defined by $x=r \,
\cos(\varphi)$ and $y=r \, \sin(\varphi)$. Outside the wire, the
right hand side of equation~(\ref{eq.A2D}) is zero. The solution
of this 2D heat equation~without source term can be expanded in a
basis of functions with a given "angular momentum" $n$. The radial
dependence of the solution is therefore a linear combination of
modified Bessel functions of the first kind $I_n$ and of the
second kind $K_n$. Thus expanding $A_{l,k}$ on this basis, we
obtain the following linear combination for the vector potential
\begin{equation}
A_{l,k}(r,\varphi) =\sum_{n=-\infty}^{n=\infty}c_{l_n}(k) \ e^{i
\, n \, \varphi } \ K_n(k \, r) \ .\label{eq.A}
\end{equation}
We retain only the modified Bessel functions of the second kind,
since the potential has to go to zero as $r$ goes to infinity. The
$c_{l_n}(k)$ coefficients are imposed by equation~(\ref{eq.A2D}),
and can be determined using the Green function of the 2D heat
equation~\cite{mathpourphysiciens}. We obtain
\begin{equation}
c_{l_n}(k)=-\frac{\mu_0}{2\pi} \int \! \! \! \int  I_n(k\,r) \
e^{-i \, n\, \varphi} \ j_{l,k}(\varphi,r) \ r \, dr \, d\varphi \
. \label{eq.defcn}
\end{equation}
Taking the curl of the vector potential and using the relations
$K_n'=-(K_{n-1}+K_{n+1})/2$ and $2\, n\,
K_n(u)/u=-K_{n-1}+K_{n+1}$, we obtain the $z$ component of the
magnetic field from equation~(\ref{eq.A})
\begin{equation}
B_{z,k} =
\begin{array}[t]{l}
\displaystyle{ -\frac{k}{2} \sum_{n=-\infty}^\infty
[c_{y_{n-1}}(k)+c_{y_{n+1}}(k)] K_{n}(k\, r)
e^{i \, n \, \varphi}} \\
- \displaystyle{ i\, \frac{k}{2}\sum_{n=-\infty}^\infty
[c_{x_{n-1}}(k)-c_{x_{n+1}}(k)] K_{n}(k \, r)e^{i \, n \, \varphi}
} \ .
\end{array}
\label{eq.devBz}
\end{equation}
This expression is valid only for $r$ larger than $r_0$, the
radius of the cylinder that just encloses the wire. At a given
distance $x$ from the wire, we expect that only fluctuations with
wavelengths larger or comparable to $x$ contribute to the magnetic
field, since fluctuations with shorter wavelengths average to
zero. Therefore we can simplify expression~(\ref{eq.devBz})
assuming we calculate the magnetic field above the center of the
wire $(y=0)$ for $x$ much larger than $r_0$. The argument of $I_n$
in equation~(\ref{eq.defcn}) is very small in the domain of
integration and we can make the approximation $I_n(k\,r) \simeq (k
\, r)^n/(2^n \, n!)$. This shows that the $c_{l_n}$ coefficients
decrease rapidly with $n$. Keeping only the dominant term of the
series in equation~(\ref{eq.devBz}), we obtain
\begin{equation}
    B_{z,k}(x) \simeq -\frac{c_{y_0}(k)}{k} \times \left[k^2 \, K_1(k\, x) \right]\ .
    \label{eq.Bgeneral_loin}
\end{equation}
We will see in the next section that the first factor of this
expression, characterizing the distortion flow, is proportional to
the power spectral density of the wire corrugation. The second
factor peaks at $k \simeq 1.3/x$ justifying the expansion.
Fluctuations with a wavelength much smaller or much larger than
$1/x$ are filtered out and do not contribute. As we approach the
wire, more and more terms have to be added in the series of
equation~(\ref{eq.devBz}) to compute the magnetic field. We
emphasize that the expressions derived in
equations~(\ref{eq.devBz}) and~(\ref{eq.Bgeneral_loin}) are
general for any distorted current flow that may arise from bulk
inhomogeneities or edge and surface corrugations.

\section{Calculation of the distorted current flow in a corrugated wire}\label{sec:j_distortion}
\begin{figure}
\includegraphics{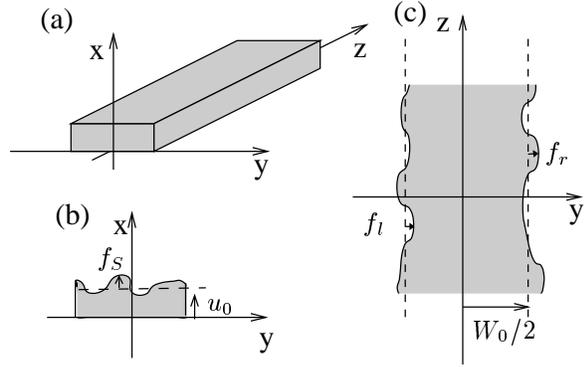}

\caption{Rectangular wire considered in this paper. The edge
roughness and the top surface roughness are illustrated in (c) and
(b) respectively.} \label{fig.wireandaxes}
\end{figure}
\begin{figure}
\resizebox{\columnwidth}{!}{%
  \includegraphics{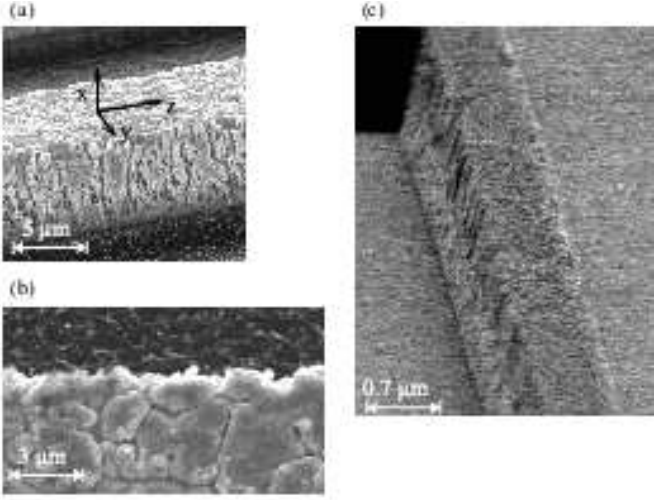}

} \caption{Scanning electron microscope images of micro fabricated
wires. Side view (a) and top view (b): electroplated gold wire of
width 50$\,\mu$m and height $4.5\,\mu$m fabricated using optical
lithography. Side view (c):
 evaporated gold wire of width and height $0.7\,\mu$m fabricated
using electron beam lithography.} \label{fig.filssem}
\end{figure}

We now turn to the calculation of the distortion in the current
flow due to wire edge and surface corrugations in order to
determine the associated $c_{l_n}$ coefficients. We suppose the
wire has a rectangular cross section of width $W_0$ and height
$u_0$ as shown in figure~(\ref{fig.wireandaxes}). Let us first
concentrate on the effect of corrugations of the wire edges,
\textit{i.e.} the borders perpendicular to the substrate (model
equivalent to~\cite{Lukin-frag2003}). Figure~(\ref{fig.filssem})
shows that, in our samples, these fluctuations are almost
independent of the $x$ coordinate both for wires deposited by
electrodeposition and by evaporation. We believe this result to be
general for wires fabricated by a lithographic process, since any
defect in the mask or in the photoresist is projected all along
the height of the wire during the fabrication process. Thus, in
the following, the function $f_{r/l}$ that describes the deviation
of the right (respectively left) wire edge from $\pm W_0/2$ is
assumed to depend only on $z$.

Conservation of charge and Ohm's law give $\nabla \vec{J}=0$ and
$\vec{J}=-\chi \, \nabla V$ where $\chi$ is the electrical
conductivity and $V$ the electrostatic potential. We will make the
approximation that $\chi$ is uniform inside the wire. In this case
$V$ satisfies the Laplace equation $\nabla^2 V=0$. As we are
interested in deviations from the mean current density $j_0=I/(u_0
\, W_0)$, we introduce the electric potential $v=V-j_0 \,z /\chi$
which is equal to zero in the absence of deviations. From what we
have said above, $v$ only depends on $y$ and $z$ and satisfies the
2D Laplace equation. The boundary conditions for the current
density on the wire edge require the current to be parallel to the
wire edge. Thus $v$ satisfies
\begin{multline}
\frac{df_{r/l}}{dz}(z) \times \left[j_0 - \chi \, \frac{\partial
v}{\partial z} (y=\pm W_0/2+f_{r/l},z)\right]=  \\
-\chi\frac{\partial v}{\partial y} (y=\pm W_0/2+f_{r/l},z) \ .
\label{eq.condilimtbord}
\end{multline}
In the following we assume the amplitude of $f_{r/l}$ to be small
enough so we can make an expansion to first order in $f_{r/l}$ of
both terms. We then obtain a linear relation between $v(\pm
W_0/2,z)$ and $f_{r/l}(z)$ which in Fourier space can be written
as
\begin{equation}\label{eq.boundary}
i \, k \, j_0 \, f_{r/l,k} =-\chi \, \frac{\partial v_k}{\partial
y} (y=\pm W_0/2) \ .
\end{equation}
The potential $v$ satisfies the 2D Laplace equation, so the $k$
component $v_k(y)$ is a linear combination of $e^{+k \, y}$ and
$e^{-k \, y}$. The two coefficients are imposed by the two
boundary conditions of equation~(\ref{eq.boundary}). To complete
the calculation of these two coefficients, we introduce the
symmetric component $f^+=(f_r+f_l)/2$ and antisymmetric component
$f^-=(f_r-f_l)/2$ of the wire edge fluctuations. Going back to the
current density, we obtain
\begin{equation}
j_{y,k}= i \, k \, j_0\left ( \frac{\cosh(k\, y)}{\cosh(k\,
W_0/2)}f^+_k +\frac{\sinh(k\, y)}{\sinh(k\, W_0/2)}f^-_k \right ).
\label{eq.jbords}
\end{equation}
We note that the symmetric part (first term) of the current
deviation is maximal near the wire edges for components with a
wave vector large compared to $1/W_0$. On the other hand, the
components with a small wave vector are constant over the width of
the wire.

We now turn to the calculation of the current distortions due to
surface corrugation. We assume the bottom surface to be flat,
since the wire is supposed to be fabricated on a flat substrate.
We denote by $f_S$ the fluctuations of the height of the wire from
its mean value $u_0$ (see figure~(\ref{fig.wireandaxes})). We
follow the same procedure as for the calculation of the effect of
the wire edge fluctuations. Now $v$ is the electrical potential
associated with the current density $j$ due to the surface
corrugation. It depends on $x$, $y$ and $z$ and satisfies the 3D
Laplace equation. To first order in $f_S$, the boundary conditions
of a current tangent to the surface of the wire are
\begin{equation}
\left\{
\begin{array}{l}
\displaystyle{ \chi \frac{\partial v}{\partial x}(x=u_0,y,z)
+j_0\frac{\partial f_S}{\partial z}(y,z) = 0}\\
\displaystyle{ \chi \frac{\partial v}{\partial x}(x=0,y,z)=0 } \ .
\end{array}
\right.
\end{equation}
and
\begin{equation}
\displaystyle{ \frac{\partial v}{\partial y}(x,y=\pm W_0/2,z)=0} .
\end{equation}
Symmetry arguments show that only the part of $f_S(y,z)$ which is
odd in $y$ contributes to the  magnetic field along $z$ in the
plane $y=0$. An even component of $f_S$ produces currents which
are symmetric under inversion with respect to the plane $y=0$.
Therefore, they cannot contribute to $B_z$ in this plane. Thus,
only the Fourier components
\begin{equation}
f_{S_{k,m}} =  \int\!\!\!\int \frac{dy \, dz}{\pi\sqrt{2 L\, W_0}}
\ e^{-i  k  z}  \sin(2  m  \pi y/W_0) \ f_{S}(y,z)
\end{equation}
contribute, where $m=1,..,\infty$. With this definition,
$f_{S_k}(y)=2\sqrt{\pi/W_0} \sum_{m=0}^{\infty}\sin(2\pi my/W_0)
f_{S_{k,m}}$. We choose this definition of the Fourier component
$f_{S_{k,m}}$ so that $\langle|f_{S_{k,m}}|^2\rangle$ is equal to
the 2-dimensional spectral density of $f_S$.
 To obtain the electric potential produced by a given component $f_{S_{k,m}}$ we use
the expansion
\begin{equation}
\sin(2  m  \pi y/W_0)=\sum_{p=0}^{\infty} \gamma_{m,p}\sin((2p+1)\pi y/W_0),
\label{eq.devsin}
\end{equation}
where
\begin{equation}
\gamma_{m,p}=\frac{-8m}{\pi}
\frac{(-1)^{m+p}}{(2(m+p)+1)(2(p-m)+1)} ,
\end{equation}
valid for $y\in[-W_0/2,W_0/2]$.
 Each $p$ Fourier component induces an electrical potential
$v_{k,m,p}$ and,  since $v$ satisfies the Laplace equation,
$v_{k,m,p}$  is a linear combination of $e^{+ \nu_p x}$ and
$e^{- \nu_p x}$ where $\nu_p=\sqrt{k^2+((2p+1)\, \pi /W_0)^2}$.
The boundary conditions on the surfaces $x=0$ and $x=u_0$
determine the coefficients and we obtain
\begin{equation}
\begin{array}{l}
-\chi\,v_{k,m,p}(x,y)=\\
\ \ i\,k\,j_0\,f_{S_{k,m}}\gamma_{m,p}\frac{\cosh(\nu_p
x )}{\sinh(\nu_p u_0)} \frac{1}{\nu_p}\sin((2p+1)\pi y/W_0).
\end{array}
\end{equation}
 With the choice of the expansion (\ref{eq.devsin}), the boundary
conditions on $y=\pm W_0/2$ are satisfied by each term.
Finally, we obtain the current density distribution
\begin{equation}
\displaystyle \left \{
\begin{array}{l}
j_{x_{k,m}}(x,y)\!\begin{array}[t]{l}
=\!\!2ik f_{S_{k,m}}j_0 \sqrt{\frac{\pi}{W_0}}\\
\displaystyle\sum_{p=0}^{\infty} \small{ \left ( \gamma_{m,p}
\frac{\sinh(\nu_p x)}{\sinh(\nu_p u_0)} \sin((2p+1)\pi y/W_0)
\right )}
\end{array}
\\
j_{y_{k,m}}(x,y)\!\begin{array}[t]{l}
=\!\!2 ik f_{S_{k,m}}j_0 \sqrt{\frac{\pi}{W_0}}\\
{\small \displaystyle\sum_{p=0}^{\infty} \left ( \gamma_{m,p}
\frac{\cosh(\nu_{p} x)}{\sinh(\nu_{p} u_0)}
\frac{(2p+1)\pi}{\nu_{p}W_0}
\cos((2p+1)\pi y/W_0) \right )}\\
\end{array}
\end{array} \right .
\label{eq.jsurface}
\end{equation}
The Fourier components $j_{l_k}$ are obtained by summing
equation~(\ref{eq.jsurface}) for $m=1,\dots,\infty$.

\section{Rough potential of a ribbon shaped wire}\label{sec:flatwire}
In this section, we combine the results of the two previous
sections to compute the $z$ component of the rough magnetic field
in the specific case of a flat rectangular wire ($u_0\ll W_0$).
This simplification enables us to obtain analytical results for a
system that is widely used in
experiments~\cite{BECmicroKetterle,vale:2004,Vuletic-VdW,JacobBEC,SchmiedmayerY,BecZimmermann}.

We do the calculation on the $x$ axis for $x>W_0/2$ (and $y=0$).
Since the wire is considered flat, we replace the volume current
density $\vec{j}$ by a surface current density $\vec{\sigma}=\int
\vec{j} \, dx$. Then we can rewrite the $c_{l_n}$ coefficients of
equation~(\ref{eq.defcn}) as
\begin{equation}
c_{l_n}(k)\!=\!-\frac{\mu_0}{2\pi} (-i)^n\!\int_0^{W_0/2}\!\!\! dy
I_n(ky) [ \sigma_{l,k}(y)+(-1)^n\sigma_{l,k}(-y)] \ .
\label{eq.cplat}
\end{equation}
We will first study the effect of wire edge fluctuations and give
universal behaviors for the magnetic field roughness. We will then
concentrate on the effect of the top surface roughness. We will
compare the relative importance of the two effects and point out
important consequences for the design of micro wires.

\subsection{Effect of wire edge roughness}
Let us first study the effect of wire edge fluctuations. Here we
derive the same results as~\cite{Lukin-frag2003} in a different
way. Note that, unlike the calculations in~\cite{Lukin-frag2003},
the calculations presented here are only valid for distances from
the wire larger than (or equal to) $W_0/2$. The expansion we use
is nevertheless useful because it converges rapidly and permits
the determination of the magnetic field roughness for any height
larger than $W_0/2$ after the calculation of a few parameters (the
$c_n$ coefficients).

 The distorted current flow has no component along the $x$ axis. The
expression of the rough magnetic field is then given by the first
sum of equation~(\ref{eq.devBz}). Taking $\varphi=0$, we can
rearrange this sum using the equalities $K_{n}(kr)=K_{-n}(kr)$ and
$c_{y_{-n}}=(-1)^n c_{y_n}$ (see equation~(\ref{eq.cplat})), we
then obtain
\begin{equation}
B_{z,k}=
-k\displaystyle\sum_{n=0}^\infty (c_{y_{2n}}(k)+c_{y_{2n+2}}(k))
K_{2n+1}(kr).
\label{eq.Bplatbord}
\end{equation}
Since only the $c_{y_n}$ with even $n$ contribute, we see from
equation~(\ref{eq.cplat}) that only the symmetric part of the
current density participates to the magnetic field. This is what
we expect from a simple symmetry argument. For the $c_{y_{2n}}$
coefficients we obtain
\begin{multline}
    c_{y_{2n}} = (-1)^{n+1} \frac{\mu_0 \, I}{\pi \, W_0} \,i \, k \, f_k^+ \times \\ \int_0^{W_0/2}
    I_{2n}(k\, y) \frac{\cosh(k \, y)}{\cosh(k \, W_0/2)} \ dy
    \label{eq.cyplat}
\end{multline}
As pointed out in the previous section, the sum over the angular
momenta $n$ in equation~(\ref{eq.Bplatbord}) converges rapidly
with $n$ if $x \gg W_0$. More precisely, the dominant term
proportional to $K_1(k\,x)$ gives the correct result within 10\%
as soon as $x>1.5 \, W_0$. As $x$ approaches $x=W_0/2$, more and
more terms contribute, for $x=W_0/2$, 20 terms have to be taken
into account to reach the same accuracy.

We now derive the response function of the magnetic field to the
wire edge fluctuation for $x>W_0/2$ which we define as
\mbox{$R(k,x)=|B_{z,k}/f^+_k|^2$}. As we already noticed in the
previous section, far away from the wire ($x\gg W_0$), only wave
vectors $k\ll 1/W_0$ are relevant. Then we can approximate the
integral in equation~(\ref{eq.cyplat}) by expanding the integrand
to zeroth order in $k \, y$. Keeping the dominant term in the
series that defines the magnetic field, we obtain the following
expression for the response function
\begin{equation}
R(k,x) \simeq  \frac{(\mu_0 I)^2}{4\pi^2 x^4} (k\, x)^4 K_1^2(k\,
x) \ .
\label{eq.Jkloin}
\end{equation}
For a given height $x$, as $k$ increases, this function increases
from zero as $k^2$, peaks at $k=1.3/x$ and finally tends
exponentially to zero. This behavior can be understood as follows.
At low wave vectors, the angle between the direction of the
distorted current flow and the $z$ axis tends to zero, thus the
contribution of these components becomes negligible. At high wave
vectors, fluctuations with a wave length shorter than the distance
to the wire average to zero.

\begin{figure}
\resizebox{\columnwidth}{!}{%
    \includegraphics{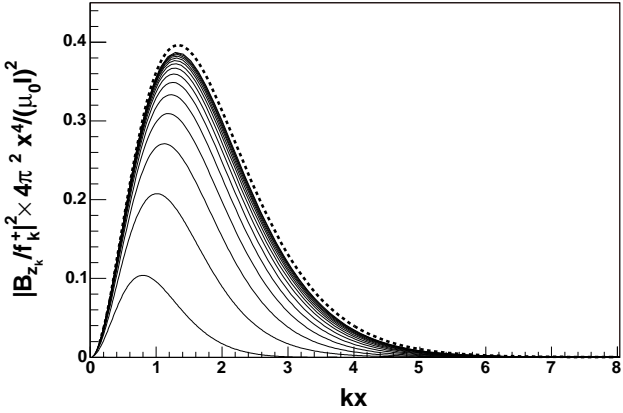}}
%~/manip/rigosite/fitbessel/pourlesfigarticle/EPJD
\caption{Response function relating the magnetic field roughness
$|B_{z_k}|^2$ to the wire edge fluctuations $|f^+_k|^2$ (see
equation~(\ref{eq.Jkloin})). Plotted is the dimensionless quantity
$|B_{z_k}/f^+_k|^2 \times 4\pi^2 x ^4/(\mu_0 I)^2$ as a function
of $k \, x$ where $x$ is the height above the center of the wire
($y=0$). The different curves correspond to different ratios
$x/W_0$ going from 0.5 to 4.7 in steps of 0.3. Small values of
$x/W_0$ correspond to lower curves. The curve corresponding to the
limit given by equation~(\ref{eq.Jkloin}) is also shown (dashed
line).} \label{fig.Jk}
\end{figure}

To check the validity of equation~(\ref{eq.Jkloin}), we plot the
dimensionless function $R(k,x)/[(\mu_0 I)^2/(4\pi^2 x^4)]$ for
different ratios $x/W_0$ in figure~(\ref{fig.Jk}).
%We find that it
%is identical to the limit function $(k\,x)^4  K_1^2(kx)$ within
%10\% as soon as $x>3.8\,W_0$.
The limit function corresponds to a configuration where the
distorted current flow is concentrated on the line $x=y=0$. For a
smaller distance from the wire, the finite width of the wire
becomes important and $R(k,x)$ differs from the
expression~(\ref{eq.Jkloin}). The amplitude is smaller and the
peak is shifted to a lower frequency. These effects are due to the
fact that as $x$ decreases, the distance to the borders of the
wire decreases less rapidly than the distance to the central part
of the wire because of the finite width of the wire. Furthermore,
because corrugations of high wave vector produce a current density
localized near the wire border, their decrease in amplitude is
more pronounced.
% Thus curiously, high wave vector components are
% slightly diminished as one approaches from the wire.

\begin{figure}
 \resizebox{\columnwidth}{!}{%
   \includegraphics{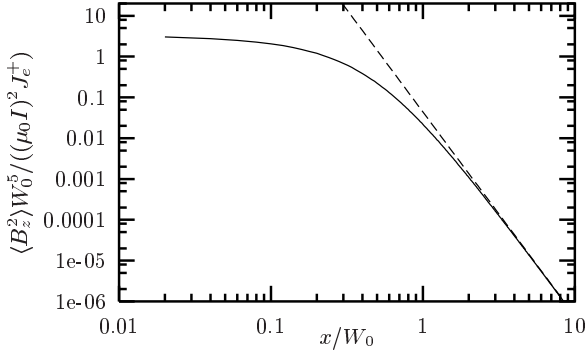}}
%\centerline{\input{Brms2.tex}}

%Cf sur backup : calculs/fragmentation ; donnees sur speck : cf papierEPJD/Brms2.plot
\caption{Magnetic field fluctuations $\langle B^2_z\rangle$ as a
function of the height above the wire ($y=0$). Plotted is the
dimensionless quantity $\langle B^2_z\rangle\,W_0^{5}/((\mu_0 I)^2
J_e^+)$, where $\langle B^2_z\rangle$ is the magnetic field
roughness and $J_e$ is the spectral density of the wire edges
assumed to be white, as a function of $x/W_0$ where $x$ is the
height above the the wire~\cite{whitenoise}. Dashed line:
$1/x^{5}$ law given by equation~(\ref{eq.Brmsloin}).}
\label{fig.Brms}
\end{figure}

Assuming a white power spectrum of the wire edge corrugations with
a spectral density $J_e^+$, we can integrate the
equation~(\ref{eq.Jkloin}) over the whole spectral
range~\cite{whitenoise}. We then find the following scaling law
for the rms fluctuations of $B_z$ with the atom-wire distance $x$:
\begin{equation}
\langle B_z^2\rangle = J_e^+\frac{(\mu_0 I)^2}{x^{5}}\times 0.044 \ .
\label{eq.Brmsloin}
\end{equation}
This expression is valid for $x \gg W_0$, the numerical factor has
been found by a numerical integration of
equation~(\ref{eq.Jkloin}). Figure~(\ref{fig.Brms}) shows that
this expression is valid within 10\% as soon as $x>2 W_0$. For
smaller distances $x$, the fluctuations of magnetic field increase
more slowly and tends to a constant. The points corresponding to
$x<W_0/2$ lie outside the range of the previous calculation and
their values have been obtained by a numerical integration   for
each $x$.
%({\it Pour vous : J'ai fait le calcul pour chaque
%composante de Fourier $k$ et j'ai intégré
%$j_k(y)K_1(k\sqrt{y^2+x^2})$.})
 Note that here $J_e^+$ is the
spectral density of $f^+$. For edges with independent
fluctuations, $J_e^+=\,J_e/2$ where $J_e$ is the spectral density
of each wire edge. The asymptotic behavior of $\langle
B^2_z\rangle$ was first derived in~\cite{Lukin-frag2003}.

\subsection{Effect of top surface corrugation}
We now consider the effect of corrugations of the top surface of
the wire.  As shown in equation~(\ref{eq.jsurface}), it induces
both a current along the $x$ and $y$ direction.
 The surface current densities obtained by integration
over $x$ have remarkably simple forms.
We find
\begin{equation}
\sigma_{y_{k,m}}=
\sigma_{y_{k,m}}^{(1)} +\sigma_{y_{k,m}}^{(2)}
\end{equation}
where
\begin{equation}
\left \{
\begin{array}{l}
\sigma_{y_{k,m}}^{(1)}=
2 i k f_{S_{k,m}} j_0  \sqrt{\frac{\pi}{W_0}}
\frac{2\pi m}{\kappa^2 W_0}\cos(\frac{2\pi m y}{W_0})\\
\sigma_{y_{k,m}}^{(2)}=
-2 i k f_{S_{k,m}} j_0  \sqrt{\frac{\pi}{W_0}}
\frac{2\pi m}{\kappa^2 W_0}
(-1)^m \frac{{\rm cosh}(ky)}{{\rm cosh}(kW_0/2)} \\
\end{array}
\right .
\label{eq.sigmaysurf}
\end{equation}
and $\kappa=\sqrt{k^2+(2m\pi/W_0)^2}$. In the calculation of
$\sigma_{x_{k,m}}$ the summation over $p$ is not analytical.
However, as we consider wires with $u_0 \ll W_0$, one can make the
approximation $({\rm cosh}(\nu_{p} u_0) -1)/{\rm sinh}(\nu_{p}
u_0) \simeq  \nu_{p} u_0$. We then obtain
\begin{equation}
\sigma_{x_{k,m}}=ikf_{S_{k,m}} j_0 \sqrt{\frac{\pi}{W_0}} u_0
\sin(2  m \pi y/W_0). \label{eq.sigmaxsurf}
\end{equation}
 Comparing equation~(\ref{eq.sigmaxsurf}) and equation~(\ref{eq.sigmaysurf}),
we see that the current density along $x$ is much smaller than the
current density along $y$ provide $\kappa \ll 1/u_0$ ({\it i.e.}
small wave vectors both along $y$ and $z$).

%as long as Fourier components $\kappa \ll 1/u_0$ are considered.
 Within our flat wire approximation, where only distances
from the wire $x\gg u_0$ are considered, this is always the case.
In the following we therefore only consider the effect of the
current density along $y$. For $x\geq W_0/2$, the rough magnetic
field is then given by equation~(\ref{eq.Bplatbord}).

\begin{figure}
 \resizebox{\columnwidth}{!}{
\includegraphics{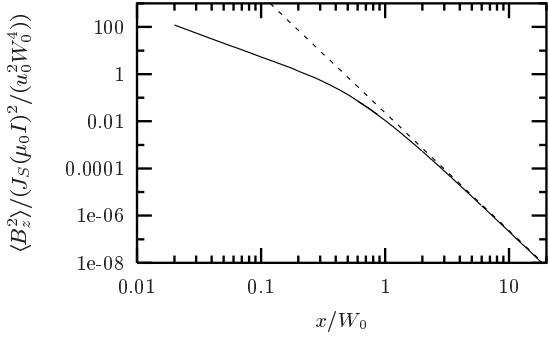}}
%/home/isabelle/manip/rugosite/fitbessel/pourlesfigarticle/EPJD/effetsurface-correct/effet-surface-derniereversion
\caption{Longitudinal magnetic field fluctuations $\langle
B^2_z\rangle$ produced by white noise top surface wire roughness
as a function of $x/W_0$ ($y=0$)~\cite{whitenoise}. Plotted is the
dimensionless quantity $\langle B^2_z\rangle /(J_S( \mu_0 I)^2
/(u_0^2 W_0^4))$.
 The dashed lines represents equation~(\ref{eq.B2loinsurf}).}
\label{fig.effetsurface}
\end{figure}

 Assuming a white power spectrum for the surface corrugation
of spectral density $J_S$, we now derive some properties of the
rough magnetic field~\cite{whitenoise}. For large distances above
the wire ($x\gg W_0$), only $k$ components much smaller than
$1/W_0$ are relevant. Then, as we have already shown, the
$c_{l_n}$ coefficients decrease rapidly with $n$ and the dominant
contribution is given by $c_{y_0}$.
 To lowest order in $ky$, $c_{y_0}$ is proportional to
the total current $\int_{-W_0/2}^{W_0/2}  \sigma_y (y)dy$. Thus,
the only contribution comes from  $\sigma_{y_{k,m}}^{(2)}$. Then,
calculations similar to those presented in the previous section
show that the contribution to $\langle B_z^2 \rangle$ of the
Fourier component $m$ of $f_{S_{k}}$ is
\begin{equation}
\langle B^2_{z,m}\rangle=
J_s\frac{W_0}{\pi u_0^2} \frac{1}{m^2}
\frac{(\mu_0 I)^2}{x^{5}}\times 0.044
\end{equation}
where $J_s$ is the 2-dimensional spectral density of $f_S$. As
expected it decreases with $m$ as the contribution of rapidly
oscillating terms averages to zero for large distances. Computing
the sum over $m>0$ gives the scaling law for the rms fluctuation
of $B_z$ due to surface corrugation with atom-wire distance $x$:
\begin{equation}
\langle B_z^2\rangle=J_s\frac{W_0}{u_0^2} \frac{\pi}{6}
\frac{(\mu_0 I)^2}{x^{5}}\times 0.044 \label{eq.B2loinsurf}
\end{equation}
 In figure~(\ref{fig.effetsurface}) this expression is compared
to numerical calculations based on equation~(\ref{eq.sigmaysurf}).
 The terms $\sigma_{y_{k,m}}^{(2)}$ contribute at least $90\%$ of
$\langle B^2_z\rangle$ as soon as $x>W_0$.

Comparing edge and surface corrugation, we see that for large
distances, both effects scale in the same way (see
equations~(\ref{eq.Brmsloin}) and~(\ref{eq.B2loinsurf})). However,
at short distances from the wire, the amplitude of the magnetic
field roughness produced by surface corrugation does not saturate.
Indeed, although the contribution of $\sigma_{y_{k,m}}^{(2)}$
saturates in the same way as the effect of wire edge fluctuations,
the contribution of $\sigma_{y_{k,m}}^{(1)}$ to the current
density diverges as one gets closer to the wire.
 Thus at small distances from the wire, we expect surface
roughness to become the dominant source of magnetic field
fluctuations.

\subsection{Consequences for micro wire traps}
 The scaling laws (\ref{eq.Brmsloin}) and (\ref{eq.B2loinsurf}) are
of major importance as they impose strong constrains in the use of
micro traps. As mentioned in section~\ref{sec:magtrap}, high
magnetic field gradients are achieved with small wires and short
distances. But as the distance to the wire decreases, the
roughness in the magnetic trapping potential increases. Imposing a
maximal roughness $\Delta B_{\rm max}$ tolerable in an experiment
therefore directly determines the maximal transverse gradient
accessible with a given realization of a micro wire.

 More precisely, as mentioned in section~\ref{sec:magtrap}, the
maximal current in a micro  wire is limited by heat dissipation:
$I_{\rm max} = \xi W_0 u_0^{1/2}$ \cite{Schmiedmayerheating}. To
analyze the scaling of the system, we consider the trap center at
a distance comparable to the wire width $x\simeq W_0$ and a wire
height $u_0$ small and constant. For a given fabrication
technology, we expect the wire roughness to be independent of the
wire dimensions $W_0$ and $u_0$ and we assume white noise spectral
densities $J_e$ and $J_S$ for the edge and top surface
corrugations~\cite{whitenoise}. Using the above expressions
 for $x$ and $I$ and equations~(\ref{eq.Brmsloin}) and
(\ref{eq.B2loinsurf}), we obtain the following scaling laws
\begin{equation}
\renewcommand{\arraystretch}{2}
\begin{array}{l}
\langle B^2_{\rm edge} \rangle = \frac{J_e \mu_0^2 \xi^2 u_0}{W_0^3} \\
\langle B^2_{\rm surf} \rangle = \frac{\pi}{6}\frac{J_S \mu_0^2
\xi^2}{W_0^2 u_0}
\end{array}
\end{equation}
for the magnetic field fluctuations induced by the edge and the
surface roughness respectively.
%Here and in the following, the
%substitution $J_e \leftrightarrow \frac{\pi}{6} \frac{J_S
%W_0}{u_0^2}$ will switch between edge and top surface effects.
 Imposing magnetic field fluctuations smaller
than $\Delta B_{\rm max}$ determines a minimal wire width
$W_{0,\rm min}$ and the maximal transverse gradient $\nabla B_{\rm
max}$. If the potential roughness is dominated by effects due to
wire edge corrugation, we find:
\begin{equation}
\begin{array}{l}
 W_{0,\rm min}= \left (
 \frac{J_e \mu_0^2 \xi^2 u_0 \times 0.044}{\Delta B_{\rm max}^2}
 \right )^{1/3} \\
 %\label{eq.maxwireedges}

\nabla B_{\rm max} = \frac{1}{2 \pi} \left ( \frac{\mu_0 \xi
\sqrt{u_0} \Delta B_{\rm max}^2}{J_e \times 0.044}
 \right )^{1/3}  .
 \label{eq.maxgradedges}
 \end{array}
\end{equation}
For a potential roughness dominated by effects due to wire top
surface corrugation, we find:
\begin{equation}
\begin{array}{l}
 W_{0,\rm min}= \left ( \frac{\pi}{6}
 \frac{J_S \mu_0^2 \xi^2 \times 0.044}{\Delta B_{\rm max}^2
 u_0}
 \right )^{1/2} \\
\nabla B_{\rm max} = \frac{1}{2 \pi} \left ( \frac{ u_0 \Delta
B_{\rm max}}{J_S \frac{\pi}{6} \times 0.044}
 \right )^{1/2} .
 \end{array}
\end{equation}
 As will be described in the following section, a micro wire
fabricated by electroplating presents an edge roughness of $J_e
\simeq 0.1\,\mu$m$^3$. Assuming a wire without top surface
roughness, a wire height of $u_0=5\,\mu$m, a typical $\xi=3\times
10^7$\,A.m$^{-3/2}$ and imposing a maximal potential roughness of
$\Delta B_{\rm max}=1$\,mG, the wire width is limited to $W_{0,\rm
min}\simeq 700\,\mu$m, the maximal gradient will be $\Delta B_{\rm
max}\simeq 0.2$\,T/cm.

\section{Probing the rough magnetic potential with cold atoms}
\label{sec:experiment}

In a previous letter~\cite{Nous-fragm}, we described measurements
of the magnetic field roughness produced by a current carrying
micro fabricated wire. The basic idea is to use the fact that the
longitudinal density $n(z)$ of atoms along the wire,
 is related to the longitudinal potential seen by the atoms
through a Boltzmann factor:
\begin{equation}
   n(z)\propto e^{-V(z)/k_{\rm B}T}.
\end{equation}
As discussed in section~\ref{sec:magtrap}, the potential $V(z)$ is
proportional to the $z$-component of the magnetic field a the
center of the trapping potential. Our typical thermal energy,
1\,$\mu$K, corresponds to a magnetic field of 15\,mG for a
$^{87}$Rb atom in the $F=2,m_{F}=2$ state. Since longitudinal
density variations of order 10\% are easily visible in our
experiment, we are sensitive to variations in the magnetic field
at the mG level.

The micro wire we used to create the magnetic potential is a
50\,$\mu$m wide electroplated gold wire of 4.5\,$\mu$m height (see
figure~(\ref{fig.filssem})). The process of micro fabrication is
the following: a silicon wafer is first covered by a 200\,nm
silicon dioxide layer using thermal oxidation. Next, seed layers
of titanium (20\,nm) and gold (200\,nm) are evaporated. The wire
pattern is imprinted on a 6\,$\mu$m thick photoresist using
optical UV lithography. Gold is electroplated between the resist
walls using the first gold layer as an electrode. The photoresist
is then removed, as well as the gold and titanium seed layers.
Finally the wire is covered with a 10\,$\mu$m layer of BCB resin
and a 200 nm thick layer of evaporated gold. The gold surface acts
as a mirror for a magneto-optical trap. The procedure for deducing
the potential roughness from images of the atomic cloud is complex
and we refer the reader to~\cite{Nous-fragm}.

\begin{figure}\label{fig.Bvsz}
\resizebox{\columnwidth}{!}{
  \includegraphics{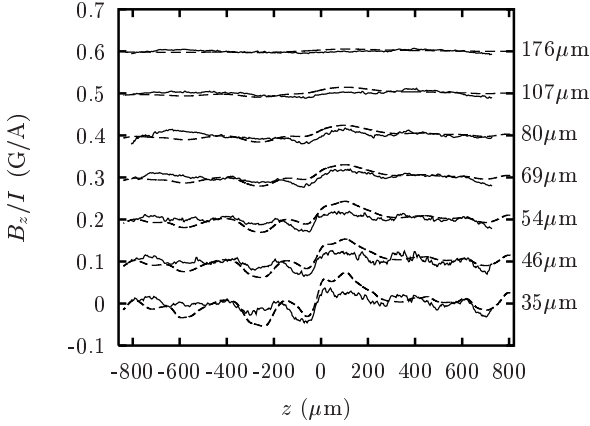}
} \caption{Rough magnetic field $B_z(z)$ normalized to the current
in the micro wire. Solid lines: magnetic field measured using cold
atomic clouds. Dashed lines: magnetic field calculated from the
measured corrugation of the edges of the wire. The different
curves have been vertically shifted by $0.1\,$G/A from each other
and heights above the wire are indicated on the right. }
\label{fig.lesBz}
\end{figure}

\begin{figure}
\centerline{\scalebox{0.7}{\includegraphics*[168pt,129pt]{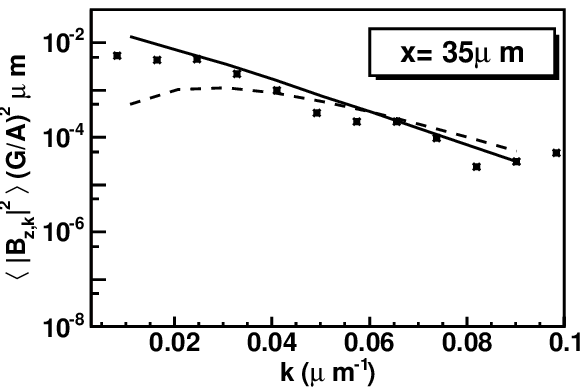}}
\scalebox{0.7}{\includegraphics*[168pt,119pt]{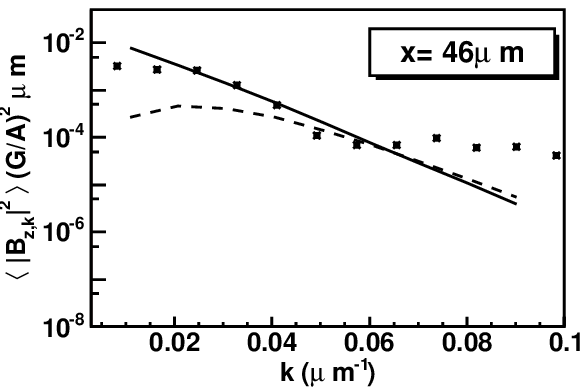}}}
\centerline{\scalebox{0.7}{\includegraphics*[168pt,129pt]{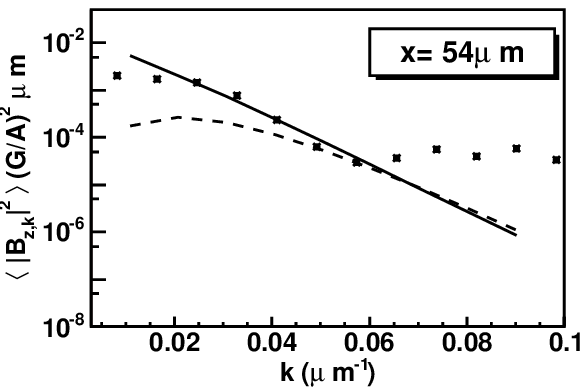}}
\scalebox{0.7}{\includegraphics*[168pt,119pt]{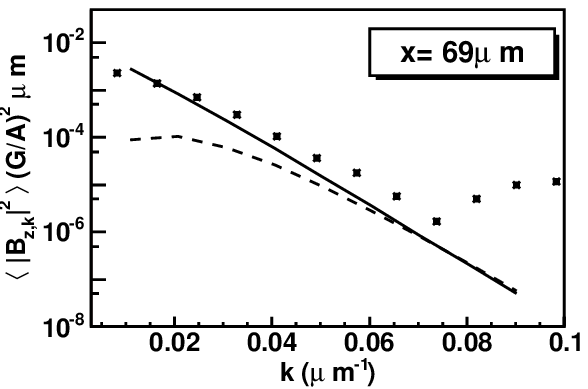}}}
\centerline{\scalebox{0.7}{\includegraphics*[168pt,129pt]{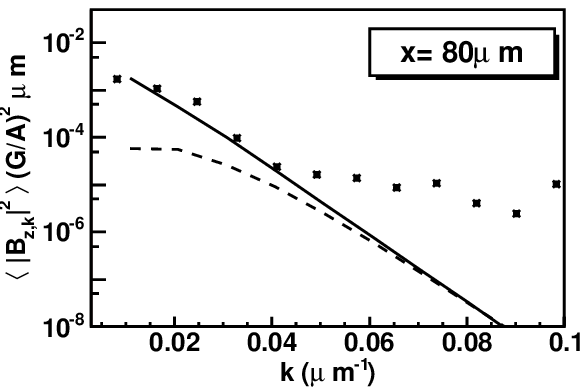}
} \scalebox{0.7}{\includegraphics*[168pt,119pt]{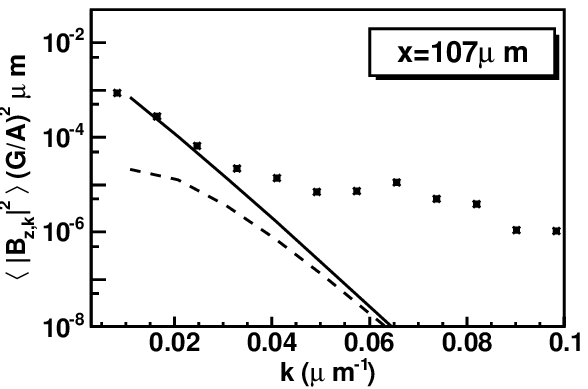}}}
\centerline{\scalebox{0.7}{\includegraphics*[168pt,129pt]{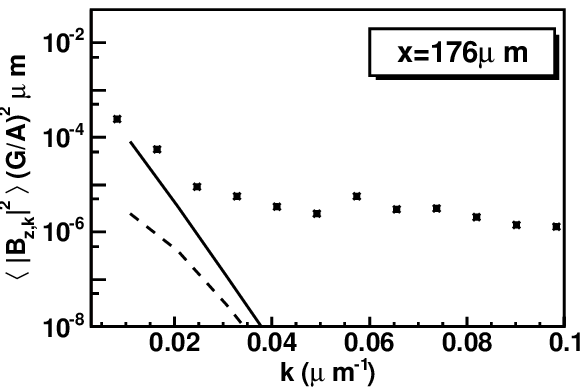}}}
%~/manip/rugosite/fitbessel/pourlesfigarticle/EPJD/compatomes
\caption{ Spectral density of the magnetic field roughness for
different heights above the wire. The points represent
experimental data. The curves result from the calculations
detailed in the text. Solid curves: expected noise due to wire
edge roughness. We used the power law fit to the spectral density
of the wire border fluctuations. Dashed curves: expected noise due
to top surface roughness.} \label{fig.compPwelch}
\end{figure}

Figure~(\ref{fig.lesBz}) shows the measured longitudinal potential
for various distances above the wire. We also show the power
spectral density of these potentials in
figure~(\ref{fig.compPwelch}). A region of 1.6\,mm along the wire
is explored by the atoms. To estimate the power spectral density
of the potential roughness we divide the total window in three
smaller windows overlapping by 50\% \cite{welch}. In each window,
the fourier transform of the potential is computed after
multiplication with a Hamming window and the estimate of the
spectral density is the average of the square of the Fourier
transforms.

In figure~(\ref{fig.compPwelch}), a flat plateau is visible at the
highest wave vectors ({\it e.g.} $k>0.07\,\mu{\rm m}^{-1}$ at
46~$\mu {\rm m}$ and $k>0.04\,\mu {\rm m}^{-1}$ at 80~$\mu {\rm
m}$). The level of this plateau depends on experimental parameters
such as the temperature and density of the atom cloud. On the
other hand the spectral density at low wave vectors, {\it i.e.} in
the region where it rises above the plateau, is independent of
these parameters. This observation leads us to conclude that while
the low wave vector part of the spectrum corresponds to a
potential seen by the atoms, the plateau at high wave vectors is
due to instrumental noise in our imaging system, such as fringes.
We expect it to vary in a complex way with temperature and atom
density. Qualitatively, smaller atom-wire distances, which are
analyzed with higher temperature clouds, should result in higher
plateaus.  This tendency is indeed observed in
figure~(\ref{fig.compPwelch}).

\begin{figure}
\centerline{\includegraphics{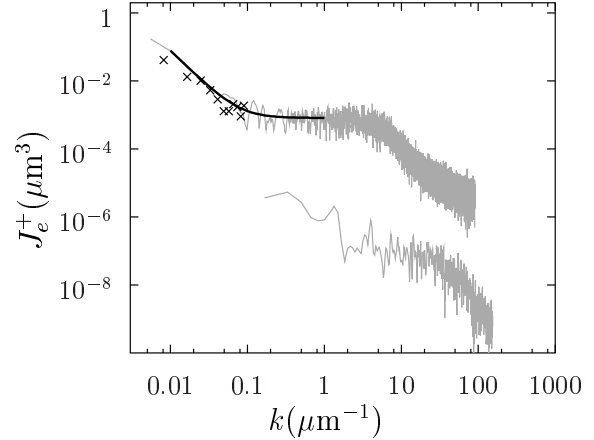}}
%~/manip/rugosite/fitbessel/pourlesfigarticle/EPJD/compatomes pour les resultats du fit
%script dans ~/manip/rugosite/papieeEPJD/
\caption{Measured spectral density of the edge roughness of the
electroplated wire (upper curve) and of the evaporated wire (lower
curve). For the electroplated wire, the spectral density of
$f^+=(f_l+f_r)/2$ is plotted. For the electroplated wire, $J_f/2$
is plotted, where $J_f$ is the spectral density of a single border
of the wire, as expected for the spectral density of $f^+$ for
independent wire border fluctuations. The crosses indicate edge
roughness for different $k$ vectors reconstructed from the decay
of the corresponding roughness measured with the atoms. The solid
line is a polynomial fit to characterize the edge roughness in the
region of interest.}

%The crosses are, for each $k$ value, the result of the fit of the
%decrease of the measured $k$ component of the magnetic field
%roughness as a function of hight above the wire.}
%The thick line on the upper
%curve is a power law fit $J=ax^{-b}+c$ which gives $b=2.19$.}
\label{fig.densspectrales}
\end{figure}

To measure the wire corrugations, we removed the atom chip from
vacuum and etched off the gold mirror and the BCB layer. We
analyzed the bare wire with scanning electron microscopy (SEM) and
with atomic force microscopy (AFM) techniques. The function $f$
describing the edge corrugation is extracted from SEM images such
as (\ref{fig.filssem}b). Rms deviations of the edges are as small
as 200\,nm, and we use a 50\,$\mu$m $\times$ 50\,$\mu$m\ field of
view in order to have a sufficient resolution. We use 66
overlapping images to reconstruct both wire edges over the whole
wire length of 2.8\,mm. We identify no correlation between the two
edges. The spectral density obtained for $f^+=(f_l+f_r)/2$ is
plotted in figure~(\ref{fig.densspectrales}). We see two
structures in the spectrum: first, we observe fluctuations with a
 correlation length of 0.2\,$\mu$m and 100\,nm rms amplitude. It
corresponds to the fluctuations seen on
figure\,(\ref{fig.filssem}b) which are probably due to the
electrodeposition process. Second, roughness with low wave vectors
is present and raises significantly the power spectral density in
the 0.01-0.1\,$\mu$m$^{-1}$ range. For the spectral range
0.01-1~$\mu m^{-1}$, the wire border fluctuations are well fitted
by a power law $J=3.2\times 10^{-6} k^{-2.15} +8.2 \times
10^{-4}\,\mu$m$^{3}$ as seen in
figure\,(\ref{fig.densspectrales}). We use this expression to
compute the spectra shown in figure~(\ref{fig.compPwelch}). As we
measured $f$ over the whole region explored by the atoms, we can
not only compare the spectral densities of the magnetic field
roughness but we also can compare the direct shape of the magnetic
field $B_z(z)$. This is done in figure~(\ref{fig.lesBz}) where the
magnetic field, computed from $f$ as described in the previous
sections, is shown by dashed lines. We note that no adjustment has
been applied to superimpose the two curves, the absolute position
of the atoms with respect to the wire is known to the 3\,$\mu$m
resolution of our imaging system.

A different approach is possible to check consistency between the
wire edge measurements and the potential roughness measured with
cold atoms. As seen in equations~(\ref{eq.Bplatbord})
and~(\ref{eq.cyplat}), $|B_{z,k}(x)|^2=J_e^+ R(k,x)$, where the
response function $R(k,x)$ does not depend on the wire edge
roughness. We compute $R(k,x)$ using the
expansion~(\ref{eq.Bplatbord}) and for a given $k$ component, we
deduce $J_e^+$ by fitting the decay of $|B_{z,k}(x)|^2$ with
height (see figure~(\ref{fig.compPwelch})). In
figure~(\ref{fig.densspectrales}), the values of $J_e^+$ obtained
by such a procedure are compared to the function $J_e^+$ measured
with the electron microscope. We find good agreement.

\begin{figure}
\centerline{
\resizebox{6cm}{!}{%
\includegraphics{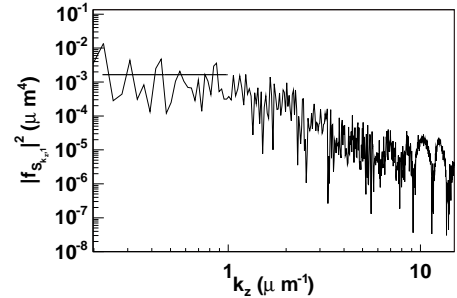}}}
%Voir /home/isabelle/manip/rugosite/fitbessel/pourlesfigarticle/EPJD/densspectralesurface
\caption{Power spectral density of the wire top surface roughness
measured with an AFM. We plot the spectral density corresponding
to the transverse mode $m=1$ ($k_y=2\pi/W_0$) which is the first
one to contribute to magnetic field roughness. The horizontal line
indicates the mean value for $k_z$ ranging from 0.2$\,\mu$m$^{-1}$
to 1$\,\mu$m$^{-1}$.} \label{fig.psdsurf}
\end{figure}

The corrugation of the top surface of the wire is measured using
an AFM and the observed power spectral density is plotted in
figure~(\ref{fig.psdsurf}).
 The spectrum is flat for wave vectors smaller than 1$\,\mu$m$^{-1}$
with a value $J_S=1.6\times 10^{-3}\,\mu$m$^4$. Unfortunately, we
were not able to obtain the spectrum for very long wave vectors.
For purposes of calculation, we shall simply assume that the
spectral density below 0.1$\,\mu$m$^{-1}$ has the same value as
between 0.1 and 1$\,\mu$m$^{-1}$. The result of this calculation
is plotted figure~(\ref{fig.compPwelch}) (dashed lines).

Our results indicate that the magnetic field roughness measured with
cold atoms is explained by  wire corrugation.
 At low wave vectors ($k<0.04\,\mu$m$^{-1}$), it seems that the magnetic
field roughness is primarily due to edge corrugations. The good
agreement between the observed field and the calculation shown in
figure~(\ref{fig.lesBz}) are the strongest evidence for this
conclusion.
 For wave vectors larger than about 0.5$\,\mu$m$^{-1}$,
the corrugations of the top surface are expected to contribute as
strongly as those of the edges. This wave vector regime however,
is not being stringently tested by our data. Since we have no data
on surface corrugation at wave vectors below 0.1$\,\mu$m$^{-1}$,
it is possible that the contribution from this effect is larger
than shown in figure~(\ref{fig.compPwelch}). The atom data in the
figure however, indicate that the surface effect is not the
dominant one although given our signal-to-noise it could be of
comparable magnitude.

Figure~(\ref{fig.filssem}b) also suggests that there might be a
grain structure in the bulk of the wires: we have no additional
information on possible current deviations due to this effect, but
the success of the model based on wire edge roughness seems to
indicate, that it is not important in our system.

\section{Improved fabrication process for micro wires}
\label{sec:evapwire} The fabrication technology described above
limits us to atom wire separations greater than several tens of
microns if we want to obtain a reasonably smooth potential. In
order to improve the quality of our wire, we turn to a different
micro fabrication process similar to~\cite{SchmiedmayerAtomChip}:
the wire structures are patterned onto an oxidized silicon wafer
using electron beam lithography. We use gold evaporation and a
standard lift-off technique to obtain 700\,nm square cross section
wires as shown in figure~(\ref{fig.filssem}c).

We extract the  wire border roughness from SEM images and the
obtained power spectral density is plotted in
figure~(\ref{fig.densspectrales}) (lower curve). In the spectral
range studied, the roughness is greatly reduced compared to the
first fabrication process. This was expected as the grain size of
evaporated gold is much smaller than of electroplated gold.
Unfortunately, we do not have a quantitative measurement of the
power spectral density in the 0.01-0.1~$\mu$m$^{-1}$ range.
Indeed, as we had to reduce the field of view to increase the
resolution, it becomes very difficult to overlap hundreds of SEM
pictures without adding spectral components due to stitching
errors. We still hope to also have reduced the wire edge roughness
in this frequency domain, as it has been demonstrated recently by
the Heidelberg group~\cite{HeidelbergPrivate} using similar wires.

Gold evaporation produces surfaces of optical quality at visible
light.
 Thus the roughness of the top surface of the evaporated wire is expected
to be much smaller than that of an electroplated wire.

\section{Conclusion}
Our goal in this paper has been to give a more detailed
description of the work which led to our conclusion that wire
corrugations can account for the magnetic field roughness
typically observed in atom chip experiments. We wish to emphasize
in this paper that great care must be taken to characterize the
roughness of a micro fabricated wire. The ratio of the rms
roughness to the wavelength of the imperfections is below
$10^{-4}$. Thus a single microscope image cannot reveal the
imperfections.

The model we use has already been suggested in
reference~\cite{Lukin-frag2003}. Here we have given more details
of the calculation as well as some physical arguments explaining
the results. We have also extended the calculation to include the
effects of corrugations of the top surface of the wire. The top
surface corrugations become increasingly important as the distance
to the wire decreases, while the effect due to wire edge roughness
saturates.

The equations~(\ref{eq.Brmsloin}) and~(\ref{eq.B2loinsurf}),
giving the behavior of the magnetic field roughness due to edge
and surface corrugation as a function of height, are important
scaling laws that one should keep in mind in the design of atom
chips. The requirements of small roughness and high transverse
confinement impose a tradeoff in choosing a wire size for a given
fabrication quality. We do not believe however that we are at the
end of our progress in improving the fabrication technology. Thus
sub-micron scale atom chips continue to hold out much promise for
the manipulation of ultra cold atoms.

We thank David Hermann for help in calculations.
This work was supported by The E.U. under grant (IST-2001-38863
and MRTN-CT-2003-505032), as well as by the DGA (03.34.033).

%\bibliographystyle{prsty}
%\bibliography{Biblio}

\begin{thebibliography}{10}
\bibitem{whitenoise}
The approximation of white spectral noise in edge or surface
roughness is used to obtain simple scaling laws and to demonstrate
the importance of the effect of wire corrugation as a whole. This
approximation is of course to be used with caution, as the noise
spectrum of micro wires strongly depends on material properties
and fabrication technique.
\end{thebibliography}
%GATHER{biblio.bib}

\end{document}